\pdfoutput=1

\documentclass[10pt,pra,aps,amssymb,amsmath,tightenlines,twocolumn]{revtex4}
\usepackage{graphicx}

\newcommand{\cC}{\ensuremath{\mathcal{C}}}
\newcommand{\cH}{\ensuremath{\mathcal{H}}}
\newcommand{\cPT}{\ensuremath{\mathcal{PT}}}
\begin{document}
\title{Winding in Non-Hermitian Systems}

\author{Stella T. Schindler}\email{sschindler@wustl.edu}
\author{Carl M. Bender}\email{cmb@wustl.edu}

\affiliation{Department of Physics, Washington University, St. Louis, MO 63130, USA}

\begin{abstract}
This paper extends the property of interlacing of the zeros of eigenfunctions in
Hermitian systems to the topological property of winding number in non-Hermitian
systems. Just as the number of nodes of each eigenfunction in a self-adjoint
Sturm-Liouville problem are well-ordered, so too are the winding numbers of each
eigenfunction of Hermitian and of unbroken $\cPT$-symmetric potentials. Varying
a system back and forth past an exceptional point changes the windings of its
eigenfunctions in a specific manner. Nonlinear, higher-dimensional, and general
non-Hermitian systems also exhibit manifestations of these characteristics. 
\end{abstract}

\maketitle

\section{Introduction}
The first observations of optical systems with balanced gain and loss
\cite{R1,R2} drew attention to $\cPT$ symmetry \cite{R3,R4}. While the concept
of $\cPT$ symmetry originated in quantum mechanics and quantum field theory, the
past decade has unearthed a trove of new experimental phenomena and potential
applications of $\cPT$ symmetry \cite{X1,X2,X3,X4}. The further discovery and
understanding of these novel behaviors requires a fundamental comprehension of
the underlying mathematical properties of $\cPT$ systems. This paper
investigates the inherent relationship between $\cPT$ symmetry and the complex
phase angle (to be defined) of an eigenfunction along a path in coordinate
space.

The spectra of non-Hermitian Hamiltonians having an {\it unbroken} $\cPT$
symmetry are entirely real \cite{R3}. Furthermore, there exists an inner product
under which these Hamiltonians possess eigenfunctions with positive norms and
exhibit unitary time evolution; thus they define physical quantum theories
\cite{R4}. These properties of $\cPT$-symmetric systems suggest the possibility
that other hallmark features of Hermitian Hamiltonians might have analogues for
$\cPT$-symmetric Hamiltonians.

A characteristic property of a Hermitian Hamiltonian is the {\it interlacing of
the zeros of its eigenfunctions}; that is, between any two consecutive {\it
nodes}, or zeros, of an eigenfunction lies exactly one node of the
next-higher-energy eigenfunction \cite{R5}. This relates directly to the
completeness of the eigenfunctions. $\cPT$-symmetric eigenfunctions also
comprise a complete set \cite{R6}; however, no $\cPT$ counterpart of Hermitian
interlacing has yet been defined. There already exists an observation of
analogous behavior of eigenfunction zeros in the complex plane; however, it is
only a numerical exploration \cite{R7}. 

$\cPT$-symmetric quantum theory uses the complex plane in order to explain and
interpret observed behaviors on the real line \cite{R4}. The $\cPT$-symmetric
extension of interlacing employs a general statement about behaviors in the
complex plane to illuminate properties in the real limit. Hermitian interlacing,
which describes relationships between countable sets of real points, is merely a
degenerate signature of a more general phenomenon in the complex plane. A slight
$\cPT$-symmetric perturbation of a Hermitian eigenfunction induces a looping
about the $x$-axis of a complex-valued, nodeless curve. The nature of the loops,
or {\it winds}, has a precise mathematical description.

We define the {\it winding number} of an eigenfunction $\psi(x)$ as the number
of times it rotates about the $x$-axis in the space $[x,\,{\rm Re}\,\psi(x),\,
{\rm Im}\,\psi(x)]$. Thus, if we take the polar decomposition of the
eigenfunction $\psi(x)=r(x)e^{i\theta(x)}$ with $r(x)$ and $\theta(x)$
real-valued, then we can express the winding number $W$ as
\begin{equation*}
W[\psi(x)]\equiv\int dx\hspace{1mm}\theta_x(x).
\end{equation*}
Throughout this paper we interpret the winding number as the overall angle in
radians traversed by a function. 

We first show that the eigenfunctions of unbroken $\cPT$-symmetric Hamiltonians
wind and that the winding numbers $W_n=W[\psi_n(x)]$ of these eigenfunctions are
distinct and well-ordered. We then explain nodal interlacing as a degenerate
signature of this well-ordered winding in Hermitian systems.

\subsection{Square-well potential}
To illustrate, we consider the square-well potential
\begin{equation*}
V(x)=\begin{cases}0&(0\leq x\leq\pi),\\ \infty&\text{otherwise},\end{cases}
\end{equation*}
with imposed boundary conditions $\psi_n(0)=\psi_n(\pi)=0$. The eigenfunctions
of this system are $\psi_n(x)=\sin(nx)$ with corresponding eigenvalues
$\lambda_n=n^2$. These eigenfunctions exhibit Hermitian interlacing because the
$n-1$ nodes of $\psi_n(x)$ lie at $x=\frac{\pi}{n},\,\frac{2\pi}{n},\,...\,,\,
\frac{(n-1)\pi}{n}$. Boundary points are not nodes \cite{R5}. 

\begin{figure}
\includegraphics[width=3in]{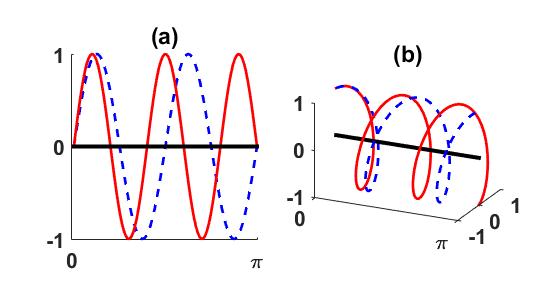}
\caption{[color online] (a) Interlacing of eigenfunctions of the Hermitian
square-well potential on the real axis. Between any two zeros of $\psi_4(x)=\sin
\,4x$ (red solid curve) lies exactly one zero of $\psi_3(x)=\sin\,3x$ (blue
dashed curve). (b) Winding of eigenfunctions of the square-well potential in the
complex plane. Note that $\psi_4(x)=e^{4ix}$ (red solid curve) and $\psi_3(x)=
e^{3ix}$ (blue dashed curve) both wind about the $x$-axis; $\psi_4(x)$ and
$\psi_3(x)$ also wind about one another (see Sec.~V).}
\label{F1}
\end{figure}

We extend these eigenfunctions into the complex plane:
\begin{equation*}
\psi_n(x)=\sin(nx)~\to~\psi_n(z)=\sin(nz).
\end{equation*}
We may now traverse a {\it complex} path to get from one boundary point to the
other. This path is parametrized as $[t,f(t)]$, where the real parameter $t$
lies in the interval $(0,\pi)$ and $f(t)$ is complex valued. We thus may write
$f(t)=f_R(t)+i\epsilon f_I(t)$, where $\epsilon$ is a real parameter and $f_R(t
)$ and $f_I(t)$ are real-valued smooth functions. Along this new path the
eigenfunctions are
\begin{eqnarray*}
\psi_n(t)&=&\sin[nf_R(t)]\,\cosh[\epsilon nf_I(t)]\nonumber\\
&&+\,i\,\cos[nf_R(t)]\,\sinh[\epsilon nf_I(t)].
\end{eqnarray*}
The hyperbolic sine and cosine functions are real-valued, positive, and
monotonically increasing on the positive-real numbers. Thus, $\psi_n(t)$ winds
in a single direction about the axis much like a helix, albeit not always at a
uniform distance from the $x$-axis. These helical eigenfunctions have distinct
and well-ordered winding numbers $W_n=n\pi$. In the limit $\epsilon\to0$ this
winding flattens out onto the real axis and reduces to the usual Hermitian
interlacing pattern (see Fig.~\ref{F1}).

An explanation of these winding numbers employs two standard theorems on
Hermitian {\it Sturm-Liouville} problems; namely, differential equations of the
general form
\begin{equation*}
-Eh(x)\psi(x)=\frac{d}{dx}\left[f(x)\frac{d\psi(x)}{dx}\right]+g(x)\psi(x).
\end{equation*}
We first note that for a rising potential $V(x)$ the Schr\"odinger equation
\begin{equation*}
i\psi_t(x,t)=-\psi_{xx}(x,t)+V(x)\psi(x,t)
\end{equation*}
possesses a countably infinite number of stationary solutions $\psi_n(x,t)=
\psi_n(x)e^{-iE_n t}$, $E_n\in\mathbb{R}$, governed by the time-independent
Schr\"odinger eigenvalue equation
\begin{equation}
\label{e1}
E\psi(x)=-\psi_{xx}(x)+V(x)\psi(x),
\end{equation}
which is in Sturm-Liouville form. The {\it Sturm-Picone Comparison Theorem} in
our case is equivalent to the statement that the eigenfunctions of a Hermitian
Hamiltonian interlace. The {\it Sturm Separation Theorem} considers any two
linearly independent solutions $u(x)$ and $v(x)$ corresponding to the same
eigenvalue of a Sturm-Liouville problem without boundary conditions imposed. The
theorem states that $u(x)$ and $v(x)$ must have the same number of nodes and
that $u(x)$ and $v(x)$ exhibit a type of interlacing. Between any two
consecutive nodes of $u(x)$ lies exactly one node of $v(x)$ and vice versa. 

For Hermitian Hamiltonians, we may always take $u(x)$ and $v(x)$ to be real
valued. The Sturm Separation Theorem implies that all extended Hermitian
eigenfunctions $\psi_n(x)=c_1u_n(x)+ic_2v_n(x)$ with $c_1,c_2\in\mathbb{R}$
exhibit winding. Furthermore, adjusting the boundary conditions to approach the
limit $c_2\to0$ corresponds to approaching the proper eigenfunctions $\psi_n(x)$
that exhibit degenerate winding number and Hermitian interlacing.

As an example, let us reconsider the square-well potential. Before we impose
boundary conditions, our extended eigenfunctions are
\begin{equation*}
\psi_n(x)=c_1\sin(nx)+ic_2\cos(nx).
\end{equation*}
By choosing $c_1=c_2=1$, we can calculate exactly the winding of
\begin{equation*}
\psi_n(x)=\sin(nx)+i\cos(nx)=ie^{-inx}
\end{equation*}
to be $n\pi$, as the eigenfunctions are perfect helices about the x-axis. Other
choices $c_1 \neq c_2$ correspond to helices with elliptical (not circular)
projections onto the plane [Re$\,\psi_n(x)$, Im$\,\psi_n(x)$] with the same
overall winding number. The eccentricity of the ellipse grows as $c_2\to0$ but
the winding number remains the same. The degenerate limiting case $c_2=0$
corresponding to the correct boundary conditions projects onto a line, which
reduces to the flat, Hermitian case of $\psi_n(x)=c_1\sin(nx)$.

The prior two explanations are fundamentally the same because the analytic
extension $\sin(nx)\to\sin(nz)$ may be expressed as $\sin(nx+iny)=\sin(nx)\cosh
(ny)+i\cos(nx)\sinh(ny)$. Curves along which ${\rm Im}\,z=y_0$ remains constant
correspond precisely to linear combinations of the linearly independent
solutions $\sin(nx)$ and $\cos(nx)$, albeit with imposed boundary condition
$\psi_n(iy_0)=\psi_n(\pi+iy_0)=i\,\sinh y_0$. In general, analytic extensions
of the eigenfunctions of Hermitian Sturm-Liouville problems always correspond to
complex sums of their two linearly independent solutions.

\subsection{Harmonic oscillator potential}
\begin{figure}
\includegraphics[width=3in]{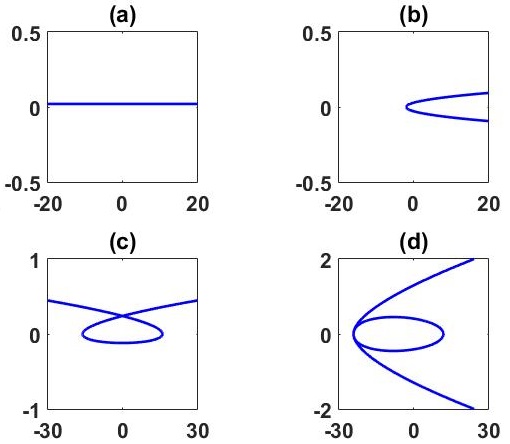}
\caption{(a)-(d) Projections of the complex Hermite polynomials $H_n(x+i\epsilon)$ onto the plane $(\text{Re}\,H_n,\,\text{Im}\,H_n)$ for fixed $\epsilon = 0.001$ and $n = 1,\,2,\,3,\,4$. Each additional half-loop traversed by a curve corresponds to an additional winding of $\pi$ traversed by the eigenfunctions of the $\cPT$-symmetric harmonic oscillator.}
\label{F2}
\end{figure}

Sturm-Liouville problems with boundary conditions at infinity also exhibit
distinct and well-ordered windings. For example, consider the
harmonic-oscillator potential $V(x)=x^2$ with eigenfunctions that vanish at $\pm
\infty$. The solutions to this problem are 
\begin{equation*}
\psi_n(x)=H_n(x)e^{-x^2/2},
\end{equation*}
where $H_n(x)$ is the $n$th Hermite polynomial. We again perform a complex
extension and find the winding numbers of the extended eigenfunctions 
\begin{equation*}
\psi_n(x+i\epsilon)=H_n(x+i\epsilon)e^{(x+i\epsilon)^2/2}.
\end{equation*}
For $\epsilon\neq0$, the system corresponds to the unbroken $\cPT$-symmetric
shifted harmonic oscillator problem $V(x) = (x+i\epsilon)^2$ \cite{R8}. The term
$e^{(x^2-\epsilon^2)/2}$ does not contribute to the winding of the
eigenfunctions. The $e^{i\epsilon x}$ term adds to each eigenfunction an
infinite number of winds about the $x$-axis independent of $n$. The distinctness
of the eigenfunction windings arises from the remaining term; that is, the
Hermite polynomials:
\begin{eqnarray*}
H_0(x+i\epsilon) &=& 1,\\
H_1(x+i\epsilon) &=& 2x+2i\epsilon,\\
H_2(x+i\epsilon) &=& 4x^2-4\epsilon^2-2+4i\epsilon x,\\
H_3(x+i\epsilon) &=& x^3-3\epsilon^2x-12x-i(12\epsilon-3\epsilon x^2+3\epsilon^3
).
\end{eqnarray*}
By integrating, we find that $H_n(x+i\epsilon)$ winds $n\pi$ times about the
$x$-axis. Therefore, the infinities cancel; that is, $W[\psi_n(x)]-W[\psi_m(x)]=
(m-n)\pi$. The eigenfunctions are well-ordered even though the windings are
infinite. We plot the projections $({\rm Re}H_n,{\rm Im}H_n)$ of the first few
Hermite polynomials in Fig.~\ref{F2}. Also, we note that in the Hermitian limit
$\epsilon\to0$, the infinite winding term $e^{i\epsilon x}$ completely
disappears from the eigenfunctions. Once more we observe the degenerate and
finite signature of winding; namely, interlacing on the real axis.

\subsection{$\cPT$-symmetric cubic potential}
Next, we consider the $\cPT$-symmetric potential $V(x)=ix^3$ on a finite domain
with eigenfunctions that are required to vanish at $x=\pm L$ \cite{R3}. This
equation is not analytically soluble, so we perform a WKB analysis of the
time-independent Schr\"odinger equation
\begin{equation*}
-\epsilon^2\psi''(x)=[E-V(x)]\psi(x)
\end{equation*}
in order to determine the behavior of high-energy eigenfunctions. We treat the
parameter $\epsilon$ as small although we will eventually set $\epsilon=1$. The
WKB approximation for $n>>1$ is
\begin{equation*}
\psi_n(x)\sim\frac{C_{\pm}}{[E_n-V(x)]^{1/4}}\exp\left[\pm\frac{i}{\epsilon}
\int^x\!\!ds\sqrt{E_n-V(s)}\right].
\end{equation*}
Setting $\psi(-L)=0$ implies that
\begin{equation}
\label{e2}
\psi_n(x)\sim\frac{C}{[E_n-V(x)]^{1/4}}\sin\left[\frac{1}{\epsilon}\int^x_{-L}
\!\!ds\sqrt{E_n-V(s)}\right].
\end{equation}
We further impose $\psi(L)=0$ and find that
\begin{equation*}
n\pi\sim\frac{1}{\epsilon}\int_{-L}^L ds\sqrt{E_n-V(s)}.
\end{equation*}
Hence, for large $E_n$, we get
\begin{equation}
\label{e3}
E_n\sim\frac{n^2\pi^2\epsilon^2}{4L^2}.
\end{equation}
Substituting (\ref{e3}) into (\ref{e2}) and setting $\epsilon=1$, we get
\begin{equation*}
\psi_n(x)\sim C\left[\frac{n^2\pi^2}{4L^2}-V(x)\right]^{-1/4}\sin\!\int^x_{-L}
\!ds\sqrt{\frac{n^2\pi^2}{4L^2}-V(s)}.
\end{equation*}
The term $C\left[n^2\pi^2/(4L^2)-V(x)\right]^{-1/4}$ contributes the same amount
of winding for each eigenfunction. Thus, we need only consider the effect of the
sine term. Substituting $V(s)=is^3$ and making a binomial approximation to the
square root, we find that
\begin{eqnarray*}
\sin\!{\int_{-L}^x\!ds\sqrt{\frac{n^2\pi^2}{4L^2}-is^3}}\!\!&\sim&\!\! \sin\!
\left(\frac{n\pi x}{2L}\right)\cosh\left[\frac{n\pi(L^4-x^4)}{16L}\right]\\
&&\hspace{-3cm}-\cos\left(\frac{n\pi x}{2L}\right)\sinh
\left[\frac{n\pi(L^4-x^4)}{16L}\right].
\end{eqnarray*}
Since $\sinh y\sim\cosh y$ for large positive $y$, we get 
\begin{equation*}
\sin{\int^x_{-L}ds\,\sqrt{\frac{n^2\pi^2}{4L^2}-is^3}}\sim
D_n\exp\left[\frac{-in\pi x}{2L}\right],
\end{equation*}
which has winding number $n\pi$ on $-L\leq x\leq L$.

For many non-Hermitian potentials, as long as $L$ remains finite, the
high-energy eigenfunctions possess finite and distinct winding numbers. If $L$
is infinite, the eigenfunctions may possess infinite but still well-ordered
winding numbers. To understand this behavior we consider the complex extensions
of the eigenfunctions $\psi_n(x)\to\psi_n(z)$. There exists a complex contour
$\cC_1$ between the turning points of $\psi_n(z)$ on which $\psi_n(z)$ is
entirely real. Furthermore, $\psi_n(z)$ has exactly $n$ nodes on this path.
However, on the other sides of the turning points there exist {\it
constant-phase contours} $\cC_2$ and $\cC_3$ from the location of the turning
point out to infinity on which the eigenfunction possesses a constant angular
argument and never vanishes \cite{R7,R9}. Thus, eigenfunctions defined along the
curve $\cC=\cC_1+\cC_2+\cC_3$ possess a Hermitian-like degenerate winding $n
\pi$. Continuously deforming $\cC$ to the real axis pulls the eigenfunction out
of the well-behaved-phase region into an oscillatory region and induces infinite
but distinct windings, like the $\cPT$-symmetric shifted harmonic oscillator
discussed in Subsec.~IB.

\subsection{Exceptional points}
To describe fully the class of $\cPT$-symmetric Hamiltonians, we insert a
parameter $\epsilon$ into a $\cPT$-symmetric potential $V(x)\to V(x;\epsilon)$.
By varying $\epsilon$, we may vary the degree of symmetry or symmetry-breaking
present in the system. By convention, we insert $\epsilon$ in such a way that
$V(x;0)$ is Hermitian. A value of $\epsilon$ at which the Hamiltonian operator
is {\it singular}, that is, at which at least one pair of eigenfunctions
possesses the same eigenvalue, is called an {\it exceptional point}. Past an
exceptional point (in a region of {\it broken} $\cPT$ symmetry) one or more
pairs of eigenvalues are complex conjugates and their corresponding
eigenfunctions satisfy $\psi_1(x)=c\,\psi_2^*(-x)$, where $c$ is some complex
constant. We define the {\it degree of symmetry breaking} of a Hamiltonian
$H(x,\epsilon_0)$ as the number of its eigenvalues that are paired or
complex-valued. 

\begin{figure}
\begin{center}
\includegraphics[width=2.5in]{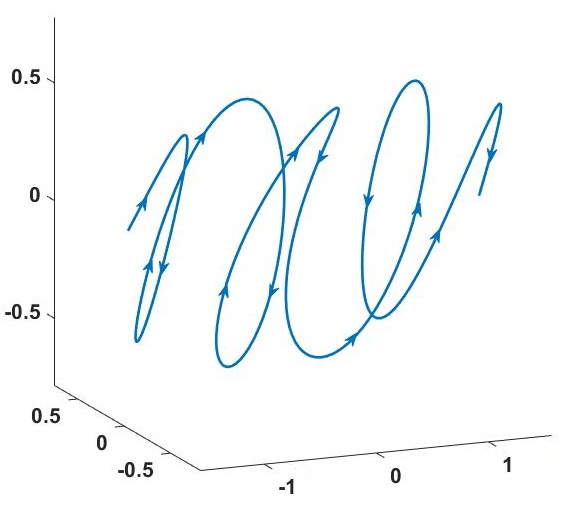}
\caption{Ninth eigenfunction of the non-Hermitian potential $V(x)=x\,\sin x+i
\epsilon\cos(3x)$ for $\epsilon=20$ plotted as the curve $[x,\,{\rm Re}\,\psi(x)
,\,{\rm Im}\,\psi(x)]$ on the interval $-\pi/2\leq x<\pi/2$. Observe that
eigenfunctions of an arbitrary non-Hermitian system need not behave as well as
eigenfunctions of a Hermitian or $\cPT$-symmetric system. As we increase
$\epsilon$ in $V(x)$ from 0 to 20, the system passes through singularities, and
thus eigenfunction windings change in an irregular manner compared to that of a
$\cPT$-symmetric system. The plotted eigenfunction has winding number {\it zero}
because the phase loops twice about the $x$-axis in one direction, turns around,
and then makes two loops in the opposite direction. Thus, in terms of phase
angle, $W[\psi_9(x)]=\int_{\pi/2}^{\pi/2}\theta_x(x)dx=0$.}
\label{F3}
\end{center}
\end{figure}

We find that distinctness and well-ordering of winding hold for both Hermitian
and unbroken $\cPT$-symmetric systems. These properties break down in a specific
manner at exceptional points. In the region of broken $\cPT$ symmetry,
well-ordering of windings still holds for the eigenfunctions with corresponding
real eigenvalue. However, extended interlacing need not hold for eigenfunctions
with complex eigenvalue. That is, a higher degree of symmetry breaking
corresponds to well-ordering by winding of fewer eigenfunctions. These
properties need not hold for general non-Hermitian systems lacking $\cPT$
symmetry (see Fig.~\ref{F3}).

Having explained how the phase angles $\theta_n(x)$ of unbroken $\cPT$-symmetric
eigenfunctions $\psi_n(x)$ depend locally on $x$, for the remainder of the
paper, we discuss the winding of eigenfunctions in a global sense, that is, as
functions of a parameter $\epsilon$. This paper is organized as follows. We
describe the dependence of eigenfunction winding on the degree of symmetry
breaking present in linear and nonlinear time-independent Schr\"odinger
equations in Sec. II. In Sec. III we demonstrate winding in a previously studied
nonlinear differential equation problem exhibiting ordered oscillations. Section
IV examines the property of well-ordered winding in higher-dimensional systems.
Finally, in Sec. V, we summarize these results and offer concluding remarks. 

\section{Parametric Dependence of Winding on $\epsilon$}
A theorem of Sturm states that the zeros of the real-valued eigenfunctions of
Hermitian Hamiltonians interlace. In Sec. I, we demonstrated that this
interlacing is a degenerate, limiting signature of the more general phenomenon
of eigenfunction winding in the complex plane. We may observe this winding by
integrating the differential equation from one boundary point to the other along
a path through the complex plane instead of along the real axis. We may also see
this winding by perturbing the boundary conditions on the eigenfunction into the
complex plane and then taking the limit as the boundary conditions approach the
real axis. Unbroken $\cPT$-symmetric Hamiltonians exhibit the same winding
behavior as Hermitian systems.

We can study the universal nature of winding in Hermitian and $\cPT$-symmetric
systems by examining parametrized classes of Hamiltonians such as the
$\cPT$-symmetric harmonic oscillator $V(x)=(x+i\epsilon)^2$ in Sec.~IB.
This example lacks phase transitions: the winding number of each eigenfunction
is uniform for all values $\epsilon\neq0$. We now turn our focus to systems with
exceptional points. 

When we vary $\epsilon$ in a region of unbroken $\cPT$ symmetry, the
eigenfunctions of $\cH(x,\epsilon)$ deform continuously, with two exceptions:
the flattening of infinite winding on an infinite domain at the point of
Hermiticity (like the harmonic oscillator) or the formation of a singularity in
the operator perhaps due to the breaking of a symmetry other than $\cPT$
symmetry. We now explain the general phenomenon of well-ordered windings
\cite{R7} as a consequence of the absence of eigenfunction nodes on the real
axis in a region of unbroken symmetry. 

Why do eigenfunctions lack nodes on the real axis? As long as the value of
$\epsilon$ does not correspond to an exceptional point, the operator $\cH(x,
\epsilon)$ is nonsingular. Thus, for a fixed nonexceptional value of $\epsilon$,
we may take a polar decomposition of any eigenfunction $\psi(x)=r(x)e^{i\theta(x
)}$. If $\psi(x)$ vanishes at some point $x=x_0$, then $r(x_0)=0$. Furthermore,
$r(x)$ is always nonnegative, so $r_x(x_0)=0$. Thus, $\psi_x(x_0)=[r(x_0)+ir_x(x
_0)\theta(x_0)]e^{i\theta(x_0)}=0$. The vanishing of $\psi(x)$ and $\psi_x(x)$
at $x=x_0$ implies that all derivatives of $\psi(x)$ vanish at $x=x_0$ because
$\psi(x)$ obeys the Schr\"odinger equation. If all derivatives of an analytic
function are zero at a point, then that function is constant. But $\psi(x)=0$
contradicts the assumption that $\psi(x)$ is an eigenfunction. Thus, $\psi(x)$
does not vanish on the real axis.

Because eigenfunctions of unbroken $\cPT$-symmetric operators are nodeless,
their windings may not exhibit any sudden discontinuities as we vary $\epsilon$.
Thus, the windings of eigenfunctions vary continuously in the region of unbroken
$\cPT$ symmetry. This in turn leads to the winding-number-based ordering of
eigenfunctions in these potentials. 

What happens at an exceptional point (where the operator is singular)? As
$\epsilon$ approaches an exceptional point, a pair of solutions begins to
coalesce. At the exceptional point, these two eigenfunctions are identical and
therefore possess the same winding number. Past the exceptional point, the
solutions $\psi_1$ and $\psi_2$ are $\cPT$ conjugates, so they still have the
same winding number: $W[\psi_1(x)]=W[\psi_2(x)]$.

The $\epsilon$ dependence of eigenfunctions with complex eigenvalue is distinct
from that of eigenfunctions with real eigenvalue. As we parametrically pass
through an exceptional point, the windings of the eigenfunctions undisturbed by
the crossing with real corresponding eigenvalue remain ordered with respect to
one another in the same manner as prior to the crossing. However, eigenfunctions
having complex eigenvalues need not respect that order.

The eigenfunctions of a non-Hermitian Hamiltonian $\cH(x;\epsilon)$ that lacks
symmetry do not necessarily exhibit any of these characteristics. Due to the
lack of consistency in singularity formation in $\cH(x;\epsilon)$ with respect
to the parameter $\epsilon$, eigenvalues may develop multiplicities or become
complex in a nonuniform fashion. Eigenfunctions may also develop nodes
unsystematically and do not shift windings in a prescribed manner past a
singular point. Thus, deformations of non-Hermitian Hamiltonians do not have to
maintain strict eigenvalue-based order or well-predicted pairings of winding
numbers.

The statements above appear to hold for all potentials, whether or not they are
periodic. However, the extra constraints that periodicity places on a system
lead to noticeable differences in the winding variations in $\epsilon$. We
present two examples to highlight these differences. We then examine how similar
properties appear even when the system is nonlinear.

\subsection{Nonperiodic potential}
\begin{figure}
\includegraphics[width=3in]{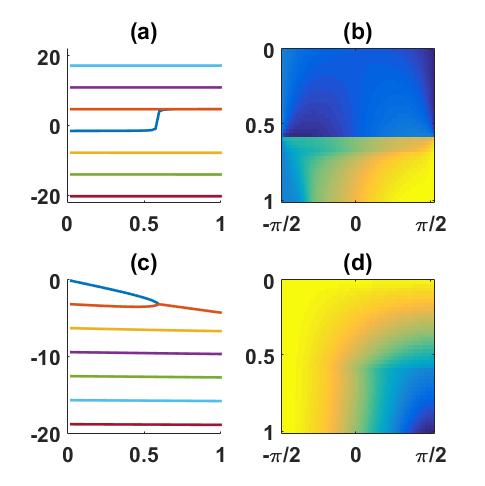}
\caption{[color online] Dependence of eigenfunction winding on the parameter
$\epsilon$ for the nonperiodic potential $V(x)=4-4i\epsilon x$ (panel a) and the
periodic potential $V(x)=4\cos^2x+4i\sin(2x)$ (panel c). The corresponding phase
angles $\theta(x,\epsilon)$ of the first eigenfunction of each potential for
a range of $(x,\epsilon)$ are shown in (b) and (d). Note the sharp jump
discontinuity in both eigenvalue magnitude and eigenfunction phase-angles
function at an exceptional point [panel (b)], compared to the continuous
dependence on $\epsilon$ for the nonperiodic potential [panel (d)].}
\label{F4}
\end{figure}

Consider first the linear Schr\"odinger eigenvalue equation (\ref{e1}) with the
nonperiodic $\cPT$-symmetric potential
\begin{equation*}
V(x)=4-4i\epsilon x
\end{equation*}
on the finite interval $[-\pi/2,\pi/2]$. We impose homogeneous boundary
conditions at the endpoints. As predicted, the eigenfunction windings deform
smoothly and continuously in the region between two exceptional points. At an
exceptional point, the winding numbers of two eigenfunctions merge. The
exceptional point is the sole point of nonsmooth variation of winding numbers;
the derivative $\frac{dW_n}{dt}$ may be different on either side of the
transition. Beyond an exceptional point, however, these winding numbers vary
smoothly once more and in tandem. That is, nonsmooth winding deformation occurs
only at an exceptional point and only for the specific eigenfunctions with
coalesced eigenvalues. We plot the phase of the eigenfunctions at each $(x,
\epsilon)$ value in Fig.~\ref{F4}. 

\subsection{Periodic potential}
The eigenfunctions of a $\cPT$-symmetric periodic potential, unlike those of a
nonperiodic potential, always possess winding numbers that are equivalent modulo
$2\pi$. This property is due to Bloch's theorem, which states that solutions to
a Schr\"odinger equation with a periodic potential take the form
\begin{equation*}
\psi_{n}(x)=u_{n}(x)e^{ikx},
\end{equation*}
where $-\pi\leq k<\pi$ is a chosen Bloch wavenumber. Thus, any change in winding
must occur as a discontinuous jump. As we approach an exceptional point the
behavior of the eigenfunctions is not immediately apparent. 

Thus, let us examine the complex phase at every point of the wave. We consider
the potential 
\begin{equation}
\label{e4}
V(x)=4\cos^2x+4i\epsilon\sin(2x),
\end{equation}
which was studied in Ref.~\cite{R10}. We first find the eigenfunctions for $k
\neq0$. Figure~\ref{F4} shows that in
the region of unbroken symmetry the eigenfunctions have winding numbers $W_{2n}
=-W_{2n+1}$. The approach towards an exceptional point is marked not only by the
formation of a sharp cusp in eigenfunction magnitude but also by the
corresponding formation of a jump discontinuity in eigenfunction phase. The
eigenfunction tries to complete a full loop about the $x$-axis within an
extremely short distance. As that length reaches zero, the system crosses an
exceptional point and the eigenfunction winding jumps. Interestingly, only one
eigenfunction of the pair exhibits this type of dependence on $\epsilon$ rather
than a mutual merging as observed in the nonperiodic case. At $\epsilon=0.5$,
all the bands go complex simultaneously at their edges $k=\pm1$. At this
parameter value, the eigenfunctions have exact solutions in terms of Bessel
functions \cite{R10}:
\begin{equation*}
u(x)=J_k(i\sqrt{\epsilon/2}\,e^{ix}).
\end{equation*}
Here, we calculate that $u(x)$ has winding number $\pi$. 

We remark that some Hamiltonian systems may have more complicated variations in
eigenfunction pairings; one example is $V(x)=\cos^2x+i\epsilon\sin^3(2x)$. The
restrictions placed on winding number by complex conjugacy pairings, coupled
with the absence of nodes, helps to explain their more nuanced behaviors. 

\subsection{Cubic nonlinearity}
\begin{figure}
\includegraphics[width=2.5in]{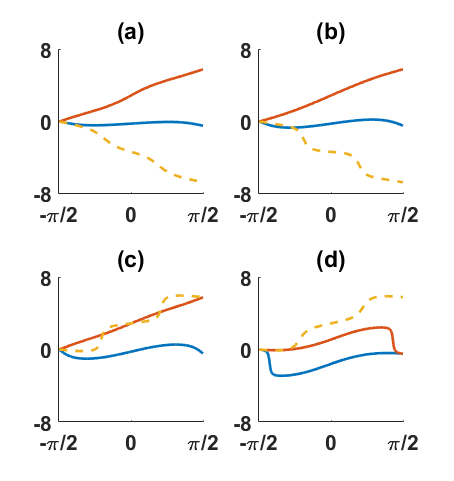}
\caption{[color online] Phase angle at every point of the first three
eigenfunctions (solid blue, solid red, dashed yellow) of the Schr\"odinger
equation with additive cubic ($|\psi|^2\psi|$) nonlinearity and potential $V(x)=
4\cos^2x+4i\epsilon\sin(2x)$ for $\epsilon=0.25,\,0.50,\,0.75,1.00$ in panels
(a)-(d). All of the eigenfunctions have real corresponding eigenvalue except for
the first and second eigenfunctions, which have become $\cPT$ conjugates [panel
(d)]. Note the approach towards a jump discontinuity in phase angle near a
singular point. The third eigenfunction is always associated with a real
eigenvalue, and the jump in its winding for $\epsilon\in(0.50,0.75)$ does not
occur because of $\cPT$ symmetry breaking.}
\label{F5}
\end{figure}

Many $\cPT$-symmetric nonlinear Schr\"odinger equations exhibit similar
characteristics to their linear counterparts. Let us first examine a
Schr\"odinger equation with an additive {\it Kerr}, or cubic ($|\psi|^2\psi$),
nonlinearity
\begin{equation*}
E\psi(x)=-\psi_{xx}(x)+V(x)\psi(x)+|\psi(x)|^2\psi,
\end{equation*}
This equation is isomorphic to the nonlinear paraxial wave equation governing
the propagation of intense light through a waveguide. For Fig.~\ref{F5}, we
analyze the extended stationary states of a periodic potential (\ref{e4})
\cite{R11,R12}. To find these states we choose a wave intensity
\begin{equation*}
P_{uc}=\int dx|\psi(x)|^2
\end{equation*}
over the unit cell. The value of $P_{uc}$ determines the reality or complexity
of the band structure \cite{R12}. Thus, adjustment of the parameters $P_{uc}$
and $E$ allows us to change the degree of symmetry breaking present at each part
of the spectrum. Singular points correspond to jumps in winding in a manner
similar to the linear case. This type of eigenfunction winding evolution occurs
regardless of the variable ($P_{uc}$, $k$, or $\epsilon$) by which we approach
an exceptional point.

If we perform the same procedure with the nonintegrable Schr\"odinger system
with quintic nonlinearity
\begin{equation*}
E\psi(x)=-\psi_{xx}(x)+V(x)\psi(x)+|\psi(x)|^4\psi(x),
\end{equation*}
using the potential in (\ref{e4}), we again find that eigenfunction windings
appear to be well ordered as in the linear and cubic-nonlinear cases.

\section{Winding Interpretation of Initial-Value Problem}
In this section we consider the nonlinear first-order {\it initial-value
problem}
\begin{equation}
y'(x)=\cos[\pi xy(x)],\hspace{5 mm}y(0)=a.
\label{e5}
\end{equation}
The solutions to this non-Sturm-Liouville problem for real $a$ on the interval
$[0,\infty)$ were described in \cite{R13}. The solutions exhibit oscillations
that look strikingly similar to the interlacing properties of Sturm-Liouville
systems. All solution curves with initial conditions in a range $a_n<a<a_{n+1}$
exhibit the same number of up-and-down oscillations. The values $...,a_{-1},a_0,
a_1,...a_n,...$ correspond to separatrix solutions with oscillation number
intermediate between the solutions with initial condition on either side. 

\subsection{Winding dependence on initial $y$ condition}
\begin{figure}
\includegraphics[width=3in]{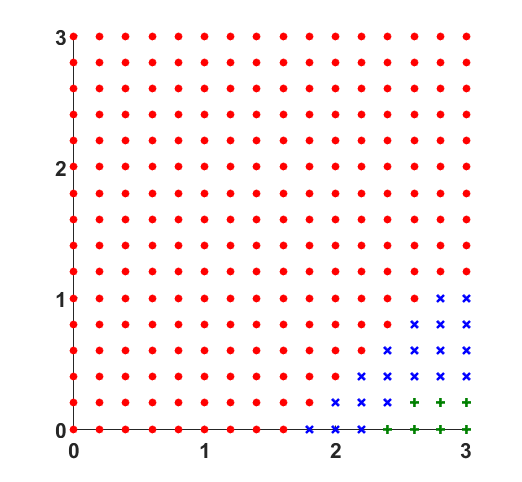}
\caption{[color online] Selected complex initial conditions and the winding
numbers to which they give rise for (\ref{e5}). A red dot indicates a winding of
$\pi$, a blue $x$ indicates a winding of $3\pi$, and a green $+$ indicates a 
winding $5\pi$. It appears that adjacent regions are separated by a curve of
initial conditions, which give rise to separatrix solutions.}
\label{F6}
\end{figure}

If we perturb the initial condition $a$ into the complex plane, $y(x)$ winds in
the space $[x,\text{Re}\,y(x),\text{Im}y(x)]$, though not about the axis $x=0$.
The winding number of $y(x)$ on the interval $[0,\infty)$ is dependent on the
initial condition $a$. Initial conditions within specific two-dimensional
regions of complex initial condition space all induce the same winding number.
These regions appear to be separated by curves of initial conditions inducing
separatrix-like behavior (see Fig.~\ref{F6}).

\subsection{Winding dependence on initial $x$ condition}
Let us take (\ref{e5}) with initial condition $y(b)=a$, with $b$ not necessarily
zero. We simplify notation by shifting the equation to get
\begin{equation}
y'(x)=\cos[\pi(x-b)(y-a)],\quad y(0)=0.
\label{e6}
\end{equation}
When we impose $y'(0)=1$, a condition that always holds for (\ref{e5}), we find
that $b$ may take on only a countably infinite number of values $b_n=2n/a$. We
denote the solution corresponding to parameter $b_n$ as $y_n(x)$.

\begin{figure}
\includegraphics[width=3in]{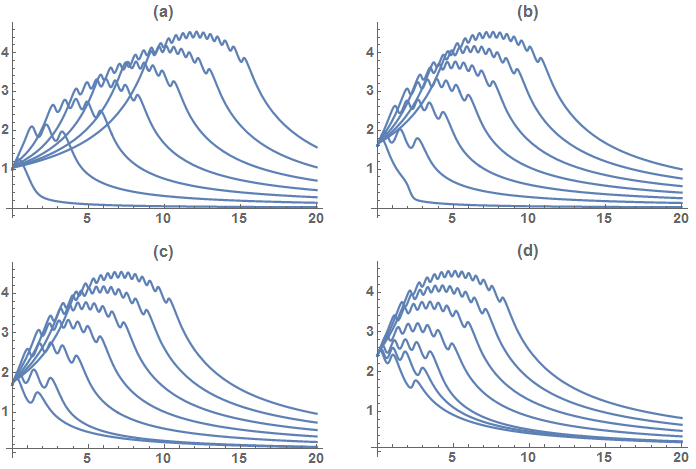}
\caption{First seven eigenfunctions of the extended cosine problem $y'(x)=\cos[
-2\pi n\psi(x)/\epsilon]$ with initial conditions $y'(0)=\epsilon$ and $y'(0)=1$
for $\epsilon=1.0,\,1.6,\,1.7,\,2.4$ in panels (a)-(d). On the interval $[0,
\infty)$, these eigenfunctions have $(4n+1)$ extrema, the degenerate signature
of winding number $(2n+1)\pi$. At each exceptional point, one or multiple
eigenfunctions each gain two local extrema, or $\pi$, of winding.}
\label{F7}
\end{figure}

We now examine a set of $y_n(x)$ for positive $n$. If $a$ is suitably small, we
find that solution curves have well-ordered winding numbers on the interval $[0
,\infty)$. As we increase $a$, the winding numbers of one or multiple $y_n(x)$
exhibit sudden shifts (see Fig.~\ref{F7}). The initial condition $b_n$ acts in a
topologically-similar manner to an eigenvalue in a Schr\"odinger equation, and
$a$ behaves similarly to the exceptional-point parameter $\epsilon$. Because
(\ref{e5}) lacks nodal interlacing, these parallels only become apparent when we
perturb these initial conditions into the complex plane and observe winding. 

Therefore, (\ref{e5}) behaves like a precise first-order differential equation
model with an exceptional-point parameter or, alternatively, like the lowest
eigenfunction of a much broader set of differential equations. In this context
the asymptotic calculations presented in Ref.~\cite{R13} mimic the calculation
of exceptional point, or operator singularity, locations. Thus, the construction
of a differential equation or its parametrization may affect the topology of
solutions. The careful addition of extra parameters or perturbation into the
complex plane may enhance our understanding of the nature of solutions and the
origin of winding phenomena.

\begin{figure}
\includegraphics[width=3in]{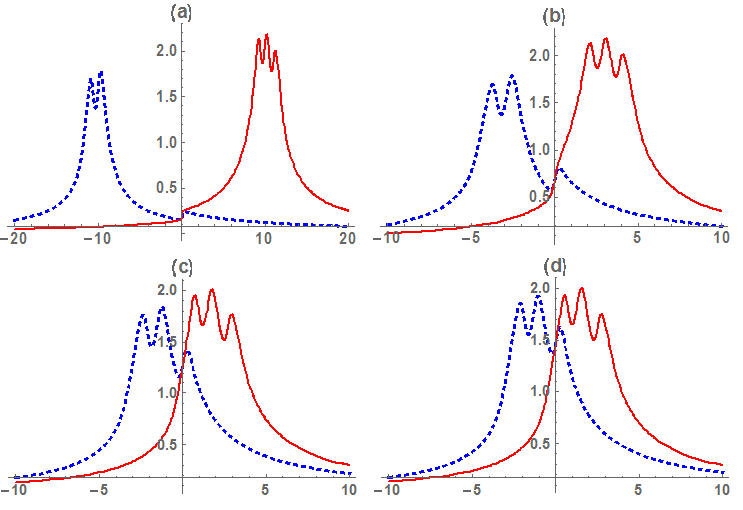}
\caption{[color online] Solutions to $y_\pm'(x)=\cos[\pi(x\pm 2\epsilon)y_{\pm}
(x)]$ for $\epsilon=0.5,\,0.7,\,1.3,\,1.5$ for panels (a)-(d). As $\epsilon$
increases, $y_{+}(x)$ (red solid curve) and $y_{-}(x)$ (blue dashed curve) pair
up and approach translations of one another along the $x$-axis, similar to the
pairing of eigenfunctions in $\cPT$-symmetric Schr\"odinger systems.}
\label{F8}
\end{figure}

Now, let us return to (\ref{e6}) and consider negative-$n$ states. On
$[0,\infty)$ for $a=0$, each solution has only one local extremum. That is, for
complex $a$, $W[y_{-n}(x)]=\pi$ for all $n$. However, as we increase $a$, the
negative-$n$ solutions develop extra oscillations (winds) two at a time, just as
the positive-$n$ solutions do. When we broaden our outlook to the full real
domain, we notice a pairing phenomenon similar to that of $\cPT$-symmetric
systems (see Fig.~\ref{F8}). While $y_n(x)$ and $y_{-n}(x)$ may differ in
winding number at $a=0$, as $a$ increases, they eventually pair off: possessing
the same winding number and approaching translations of one another along the
$x$-axis.

These properties beg whether there exist other general mathematical features to
distinguish negative-$n$ solutions from one another prior to the crossing of
singular points in the parameter $a$. 

\section{Multidimensional Systems}
Winding is not unique to the eigenfunctions of one-dimensional systems. The
solutions to a complex N-dimensional Schr\"odinger equation
\begin{equation*}
E\psi(\vec{x})=-\nabla^2\psi(\vec{x})+V(\vec{x})\psi(\vec{x})
\end{equation*}
are N-dimensional manifolds looping about the $x_1$-plane in an
$(N+2)$-dimensional space. As in Sec. II we may show that $\cPT$-symmetric
potentials $H(\vec{x};\epsilon)$ may not possess nodes except at an exceptional
point. Thus, we may begin to visualize an $N$-dimensional extension of winding
based on how many times the manifold wraps around its domain, that is, the {\it
topological degree} of the mapping. To see this, we consider two examples.

\subsection{Square-well potential}
We start with the $n$-dimensional square-well potential $V(\vec{x})=0$. This
equation has separated solutions $\psi_{a_1 a_2...a_n}(\vec{x})=\psi_{a_1}(x_1)
...\psi_{a_n}(x_n)$, where $n$ integer parameters $a_k$ are necessary to
describe the eigenfunction. The $\psi_{a_i}(x_i)$ terms are each solutions to
the one-dimensional version of this system seen in Sec.~I, $\psi_a(x)=\sin(ax)$.
If we move the boundary conditions into the complex plane, the eigenfunctions
begin to exhibit monotonically increasing phase angles in each positive $x_i$
direction. Just as in Sec.~I, we can perturb the equation so that the
eigenfunctions are of the form
\begin{equation*}
\psi_{a_1...a_n}(\vec{x}) = e^{ia_1x_1}e^{ia_2x_2}...e^{ia_nx_n}.
\end{equation*}
This perturbation does not appear to affect the phase angle at any point $\vec{x
}$, and thus the overall $n$-dimensional winding number is the same as in the
limit of vanishing boundary conditions. This function increases in phase angle
by $(a_1+...+a_n)\pi$ from the top left corner to the bottom right corner of the
domain.

\subsection{Harmonic oscillator}
\begin{figure}
\includegraphics[width=3in]{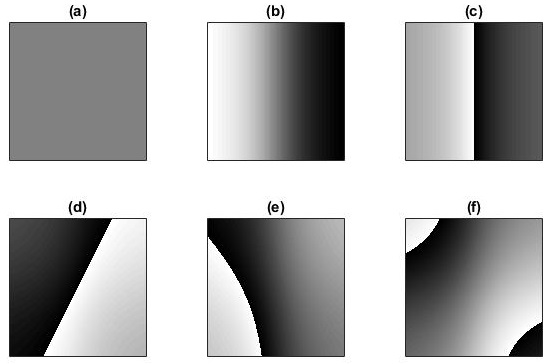}
\caption{Two-dimensional plot of the phase angle $\theta(\vec{x})$ modulo $2\pi$
of $\psi(\vec{x})=r(\vec{x})e^{i\theta(\vec{x})}$ for the two-dimensional
harmonic oscillator with both $\vec{x}$ components on the interval $[-20,20]$,
for $(m,n)=(0,0),\,(1,0),\,(2,0),\,(1,1),\,(2,1),\,(2,2)$ in panels (a)-(f).
Lighter color corresponds to increasing phase angle from $-\pi$ to $\pi$. In
general, eigenfunctions of an $N$-dimensional $\cPT$-symmetric Schr\"odinger
equation are $N$-dimensional manifolds in an $(N+2)$-dimensional space ($N$
spatial dimensions plus two extra dimensions for the real and imaginary parts of
the value $\psi(\vec{x})$ at a point $\vec{x}$).}
\label{F9}
\end{figure}

Like the square-well potential, the $n$-dimensional harmonic oscillator has
separated eigenfunction solutions
\begin{equation*}
\psi_{a_1...a_n}=e^{x_1^2+...+x_n^2}H_{a_1}(x_1)...H_{a_n}(x_n)
\end{equation*}
defined by $n$ parameters $a_i$. When we perturb $x_i\to x_i+i\epsilon_i$, the
eigenfunctions exhibit infinite winding due to the term exp[$\sum_n i\epsilon_n
x_n$]. However, this term contributes the same infinite amount of winding for
any choice $a_1,...,a_n$. The Hermite polynomials add distinctness and order of
eigenfunction windings in the system (see Fig.~\ref{F9}). Multidimensional
systems with exceptional points will be explored in future work.

\section{Conclusions and Outlook}
In this paper we have generalized the Sturm-Liouville interlacing property to
ordered winding in non-Hermitian systems. The nature of these winds and their
magnitude relative to one another may be studied by inserting a parameter into
the system and adjusting its value past singular points. In particular, the
eigenfunctions of Hermitian and unbroken $\cPT$-symmetric systems have
well-ordered winding numbers that pair up as the system passes through an
exceptional point. 

We have seen that similar descriptions apply to some nonlinear and
higher-dimensional systems. Notably, winding characterizes wave propagation in
certain time-dependent systems \cite{R14} such as the stable oscillatory
solutions of the nonlinear paraxial wave equation in \cite{R12}. Additionally,
the solutions to initial-value problems, such as (\ref{e5}) and the Painlev\'e
transcendents, exhibit real-axis oscillatory properties that translate to
winding in the complex plane \cite{R15,R16}. It is of interest to find out how
winding manifests itself in similar systems. 

Whether there exist more general statements than interlacing on the solutions of
$\cPT$-symmetric and general non-Hermitian systems remains open. One interesting
question to examine might lie in the mathematical theories of \textit{braids}
and \textit{knots} \cite{R17}, which bear resemblance to the eigenfunctions of
systems with identical boundary conditions at the endpoints. For example, it is
readily apparent that the extended eigenfunctions $\psi_n(x)=e^{inx}$ of the
Hermitian square-well potential discussed in Sec. I exhibit braiding not only
about the $x$-axis, but also about one another. That is, the $n$th and $(n+1)$st
eigenfunctions intertwine about each other exactly $n$ times [see, for example,
Fig.~\ref{F1}(b)]. We cannot deform the eigenfunctions so that they become
unwound without moving their endpoints. It follows that symmetry-breaking
exceptional points demarcate a change in braiding of the eigenfunctions. If
generally true, such an interlacing extension might help further unravel
solution behaviors.

To gain a fundamental understanding of the nature of both Hermitian and
$\cPT$-symmetric systems, it is often necessary to broaden our perspective into
the complex plane. We have extended Hermitian interlacing properties to
non-Hermitian $\cPT$-symmetric Sturm-Liouville problems. We hypothesize that
similar analyses may provide insights into nonlinear and higher-dimensional
systems described by broader classes of ordinary- and partial-differential
equations. These have potential ramifications for understanding systems in
applied disciplines.\\

STS thanks H.~Herzig Sheinfux, Y. Lumer, and M. Segev for helpful discussions
and assistance. Figures were generated using MATLAB and Mathematica.

\end{document}